\documentclass{PoS}

\usepackage{braket}

\title{Colour fields in gauge invariant quenched SU(3) Lattice QCD}

\ShortTitle{Colour fields in SU(3) quenched QCD}

\author{\speaker{Pedro Bicudo} \\ 
        CFTP, Instituto Superior T\'ecnico\\
        E-mail: \email{bicudo@ist.utl.pt}}

\author{Marco Cardoso\\
        CFTP, Instituto Superior T\'ecnico\\
        E-mail: \email{mjdcc@cftp.ist.utl.pt}}

\author{Nuno Cardoso\\
        CFTP, Instituto Superior T\'ecnico\\
        E-mail: \email{nunocardoso@cftp.ist.utl.pt}}

\abstract{
The colour fields, created by static sources belonging to different SU(3) representations,
from the $\mathbf{3}$ to the $\mathbf{27}$,
are computed in quenched SU(3) lattice QCD, in a $24^3\times 48$ lattice at $\beta=6.2$
and $a=0.07261(85)\,fm$.
We utilize the technique of generalized Wilson Loops to localize the sources, correlated with plaquettes
to measure the respective colour fields.
We investigate the Casimir scaling of the fields, measured in the static potentials by Bali.
We also study the coherence length, comparing with the dual Ginzburg-Landau approach.
With the penetration and coherence lengths we determined the Ginzburg-Landau dimensionless parameter,
this result is consistent with a type II superconductor picture, and with an effective dual gluon mass of
$0.905\pm0.163$ GeV.
}

\FullConference{The XXVIII International Symposium on Lattice Field Theory, Lattice2010\\
		June 14-19, 2010\\
		Villasimius, Italy}

\begin{document}

\section{Introduction}

Here, we compute in Lattice QCD the colour fields produced by static colour sources, with the aim to study the dual superconductor analogy of QCD. Detailed studies of the field profile of the tubes, including the string merging/separation, the Casimir scaling, and the penetration in the vacuum,
are presented.

In section 2, we review the calculation of chromo-fields in the lattice, using the Wilson Loop.
Then, in section 3 the numerical results are given for different cases. Namely,
for flux tube profile of the static $gq\overline{q}$ in a U shape geometry. For the Casimir coefficientes of the different
representations of $SU(3)$ and for the dual superconductor parameters of QCD.
We conclude in section 4.

\section{The Wilson Loops and Colour Fields}

\begin{figure}[t!] \begin{centering}
    \includegraphics[width=0.5\linewidth]{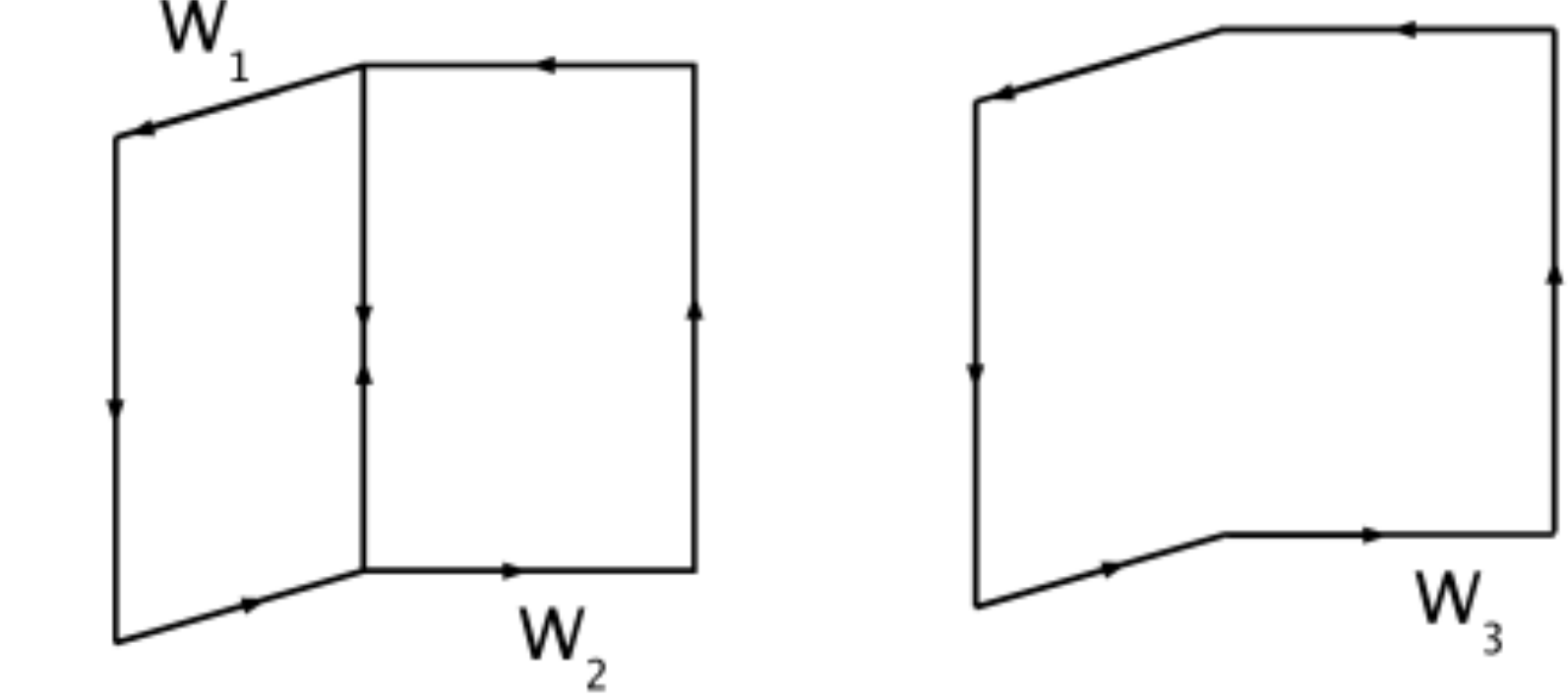}
    \hspace{1cm}
    \includegraphics[width=0.3\linewidth]{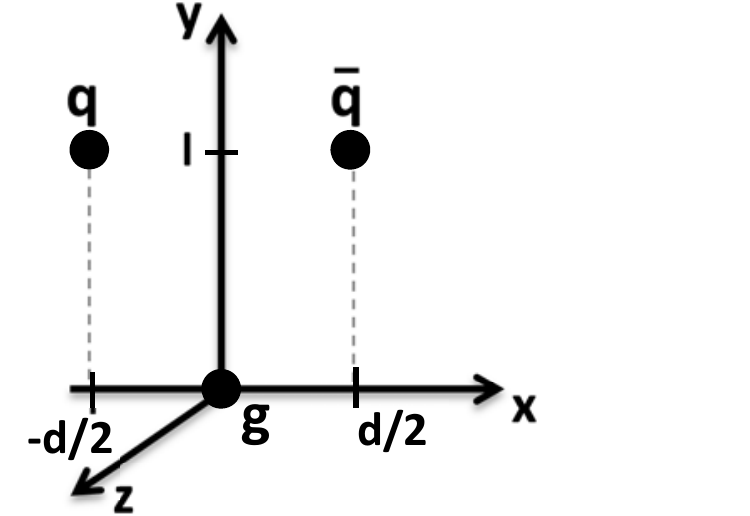}
    \caption{(left) Simple Wilson loops that make the $gq\overline{q}$ Wilson loop.
(right) Our U shape gluon-quark-antiquark geometry.}
    \label{shape +  loop}
\end{centering} \end{figure}

The Wilson loop for the static hybrid $gq\overline{q}$ was deducted in
\cite{Bicudo:2007xp,Cardoso:2007dc,Cardoso:2009qt,Cardoso:2009kz},
therefore we only present the fundamental expressions. The Wilson loop for this system is given by
\begin{equation}
    W_{gq\overline{q}}=W_1 W_2 - \frac{1}{3}W_3
\label{wloopqqg}
\end{equation}
where $W_1$, $W_2$ and $W_3$ are the simple Wilson loops shown in Fig. \ref{shape +  loop}.

We also study the interaction between two charges of different representations of $SU(3)$
in a colour singlet. So for each different $SU(3)$ representation we will have a different Wilson Loop.
The different Wilson loops are 
\cite{Bali:2000un},
\begin{eqnarray}
  W_3 &=& \mbox{tr} U  \\
  W_8 &=& |W_3|^2 - 1 \\
  W_6 &=& \frac{1}{2} \big( (\mbox{tr} U)^2 + \mbox{tr} U^2 \big) \\
  W_{15a} &=& \mbox{tr} U W_6 - \mbox{tr} U \\
  W_{10} &=& \frac{1}{6} \big( (\mbox{tr} U)^3 + 3 \mbox{tr} U (\mbox{tr} U)^2 + 2 \mbox{tr} U^3 \big) \\
  W_{24} &=& \mbox{tr} U W_{10} - W_6 \\
  W_{27} &=& |W_6|^2 - |W_3|^2 \\
  W_{15s} &=& \frac{1}{24} \big( (\mbox{tr} U)^4 + 6 (\mbox{tr} U)^2 \mbox{tr} U^2 + 3 (\mbox{tr} U^2)^2
    + 8 \mbox{tr} U \mbox{tr} U^3 + 6 \mbox{tr} U^4 \big)
\end{eqnarray}

In order to improve the signal to noise ratio of the Wilson loop, we use APE smearing, \cite{Cardoso:2009kz},
with $w = 0.2$ and iterate this procedure 25 times in the spatial direction. To achieve better accuracy in the flux tube,
we apply the  hypercubic blocking (HYP) in time direction, \cite{Hasenfratz:2001hp}, with $\alpha_1=0.75$, $\alpha_2=0.6$
and $\alpha_3=0.3$.

\begin{figure}[t!]
\begin{centering}
\includegraphics[height=0.3\linewidth]{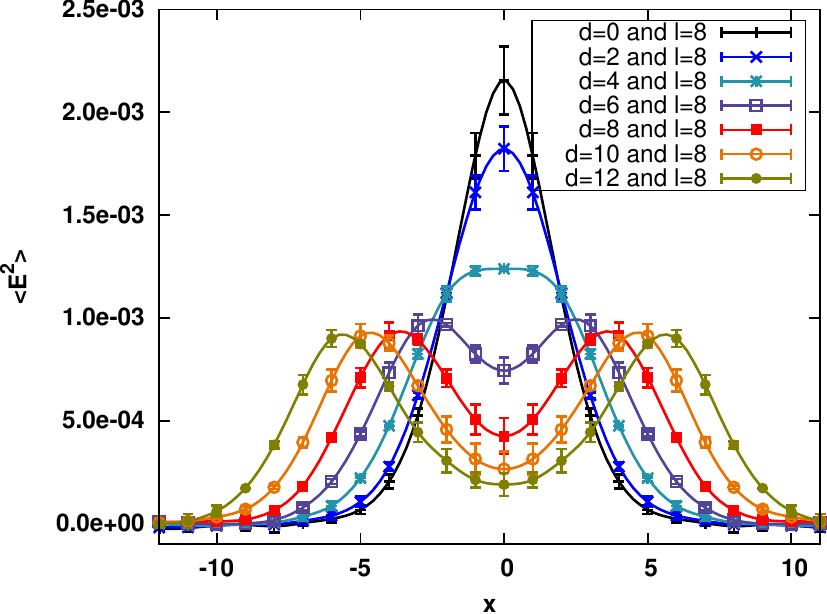}
\hspace{0.07\linewidth}
\includegraphics[height=0.3\linewidth]{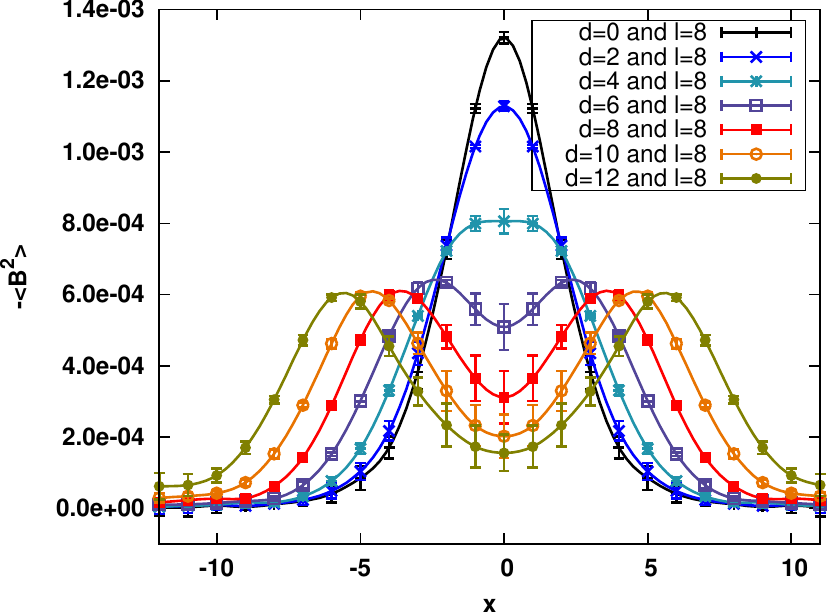}
\newline
\includegraphics[height=0.3\linewidth]{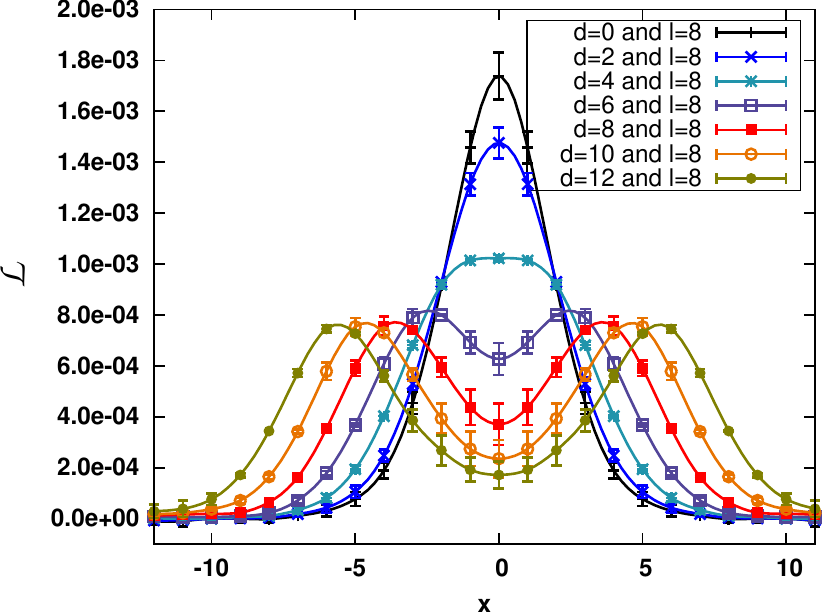}
\hspace{0.07\linewidth}
\includegraphics[height=0.3\linewidth]{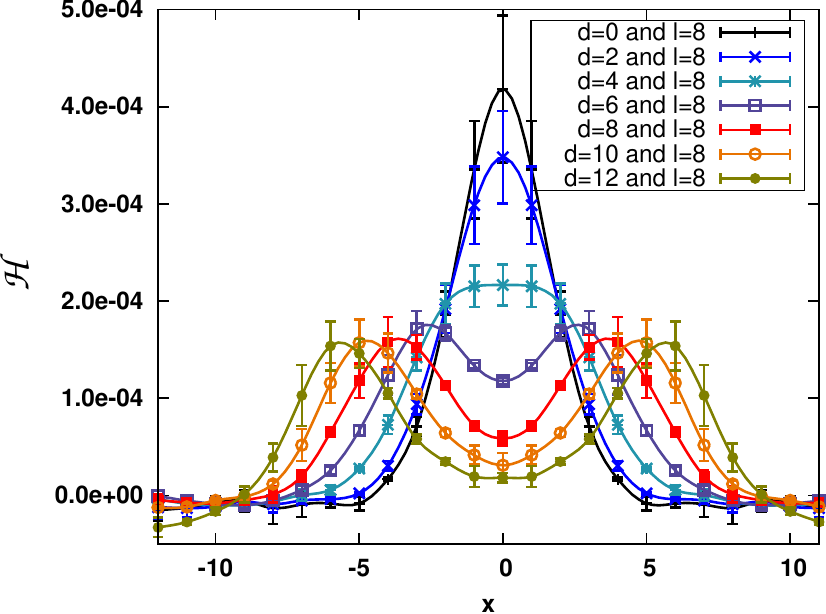}
\end{centering}
\caption{Results for the U geometry at $y=4$ and $z=0$.}
\label{qqg_U_Sim_profile}
\end{figure}

The chromoelectric and chromomagnetic fields are given by,
\begin{eqnarray}
    \Braket{E^2_i} & = & \Braket{P_{0i}}-\frac{\Braket{W\,P_{0i}}}{\Braket{W}}\\
    \Braket{B^2_i} & = & \frac{\Braket{W\,P_{jk}}}{\Braket{W}}-\Braket{P_{jk}}
\end{eqnarray}
where the $jk$ indices of the plaquette, $P$, complement the index $i$ of the chromomagnetic field. Notice that we only apply APE and HYP to the Wilson loop
and not to the plaquette $P$.

The energy ($\mathcal{H}$) and lagrangian ($\mathcal{L}$) densities are given by
\begin{eqnarray}
    \mathcal{H} & = & \frac{1}{2}\left( \Braket{E^2} + \Braket{B^2}\right)\\
    \mathcal{L} & = & \frac{1}{2}\left( \Braket{E^2} - \Braket{B^2}\right)\,.
\end{eqnarray}

We will also try to fit the results to the solutions of the Ginzburg Landau equations of
superconductivity, which in Lorentz-Heaviside units are given by
\begin{eqnarray}
    \frac{\hbar^2}{2 m} ( \nabla - i \frac{q}{\hbar c} \mathbf{A} ) \Psi + a \Psi + b |\Psi|^2 \Psi &=& 0 \\
    \nabla \times \nabla \times \mathbf{A} &=& \frac{q \hbar}{ 2 i m c } ( \Psi^* \nabla \Psi - \Psi \nabla \Psi^*
    - \frac{2 i q}{\hbar c} |\Psi|^2 \mathbf{A} ) \, .
\end{eqnarray}
This equations define two length scales, the coherence length $\xi$
and the penetration length $\lambda$,

\begin{eqnarray}
    \xi = \sqrt{\frac{\hbar^2}{2 m a}} \qquad \lambda = \sqrt{\frac{ m c^2 b }{ q^2 a }} \, .
\end{eqnarray}

\section{Results}

Here we present the results of our simulations with 286 $24^3 \times 48$, $\beta = 6.2$ quenched configurations generated
with the version 6 of the MILC code \cite{MILC}, via a combination of Cabbibo-Mariani and overrelaxed updates.
The results are presented in lattice spacing units of $a$, with $a=0.07261(85)\,fm$ or $a^{-1}=2718\,\pm\, 32\,MeV$.

In this work a geometries for the hybrid system, $gq\overline{q}$, is investigated: a U shaped, defined in Fig. \ref{shape +  loop}. 
We also investigate the case of two static colour charges only,
but for different SU(3) representations, staring by the wuark-antiquark
and gluon-gluon system, but we also study higher representations up to the
$15c$ .

The use of the APE (in space) and HYP (in time) smearing allows us to have better results for the flux tube,
while suppressing the fields near the sources.

\subsection{U profiles}

\begin{figure}[t!]
    \includegraphics[width=0.4\linewidth]{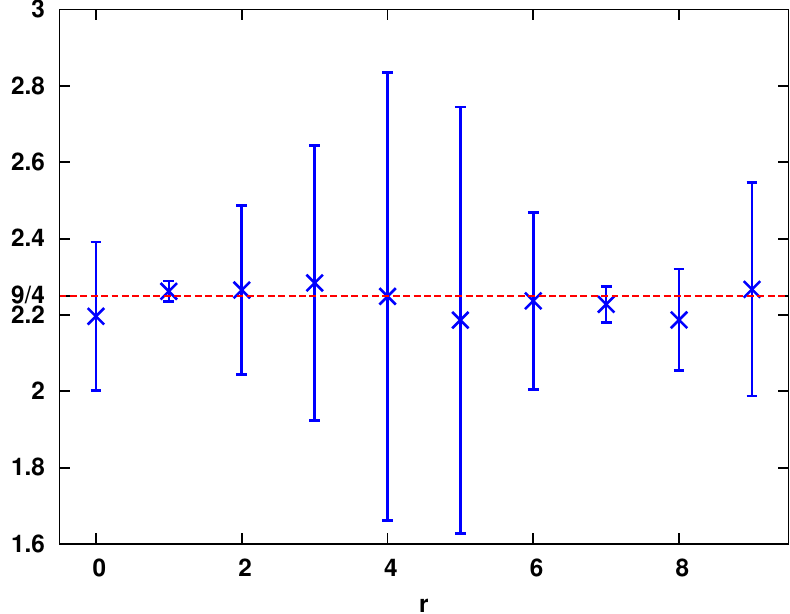}
    \includegraphics[width=0.4\linewidth]{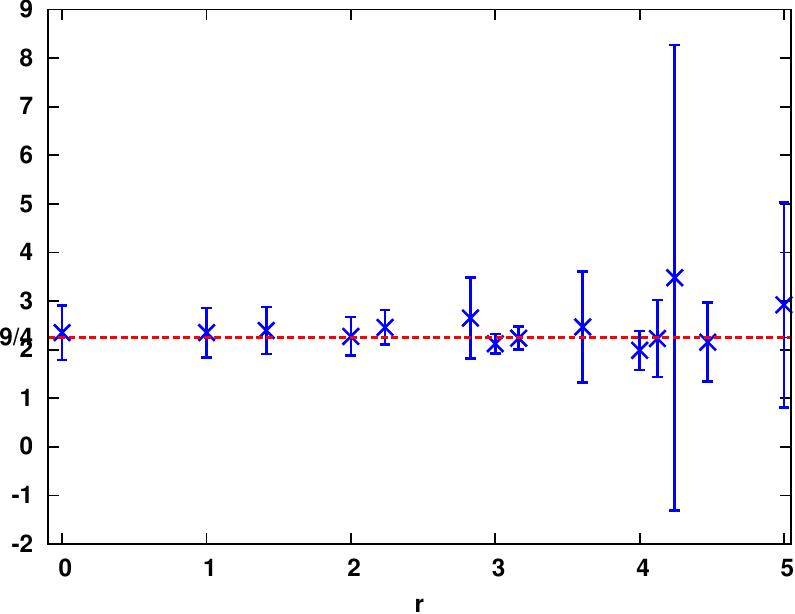}
    \caption{ On the left, results for  $r=y$ at $x=0$ and $z=0$. On the right, for
    $r=\sqrt{x^2+z^2}$ at the mediator plane.}
    \label{casimir1}
\end{figure}

In Fig. \ref{qqg_U_Sim_profile} we present the profiles for the U geometry for $l=8$ and $d$ between 0 and 16 at $y=4$.
We can see the stretching and partial splitting of the flux tube in the equatorial plane ($y = 4$)
between the quark and antiquark.
For $d \leq 4$ the separation between the two flux tube is not visible. This is due to the overlap of the tails of the flux tube profile, i. e. to the width of the flux tubes.  For larger separations the two flux tubes maxima are separated, while the tails of the flux
tubes contribute to a non zero field at $x=0$.

\subsection{Casimir scaling}

We measure the quotient between the energy density of the two gluon glueball system and of the meson system, in the
mediator plane between the two particles ($x = 0$).
The results are shown in Fig. \ref{casimir1}.

We make a constant fit to the data in Fig. \ref{casimir1}, the result for  $r=y$ at $x=z=0$ is
$2.25 \, \pm \, 0.02$ and for $r=(x,z)$ at $y=4$ is $ 2.24 \, \pm \,0.06$.
As can be seen, these results are consistent with Casimir scaling, with a factor of $9/4$ between the energy
density in the glueball and in the meson. This corresponds to the formation of an adjoint string.

\begin{figure}[t!]
    \includegraphics[width=0.4\linewidth]{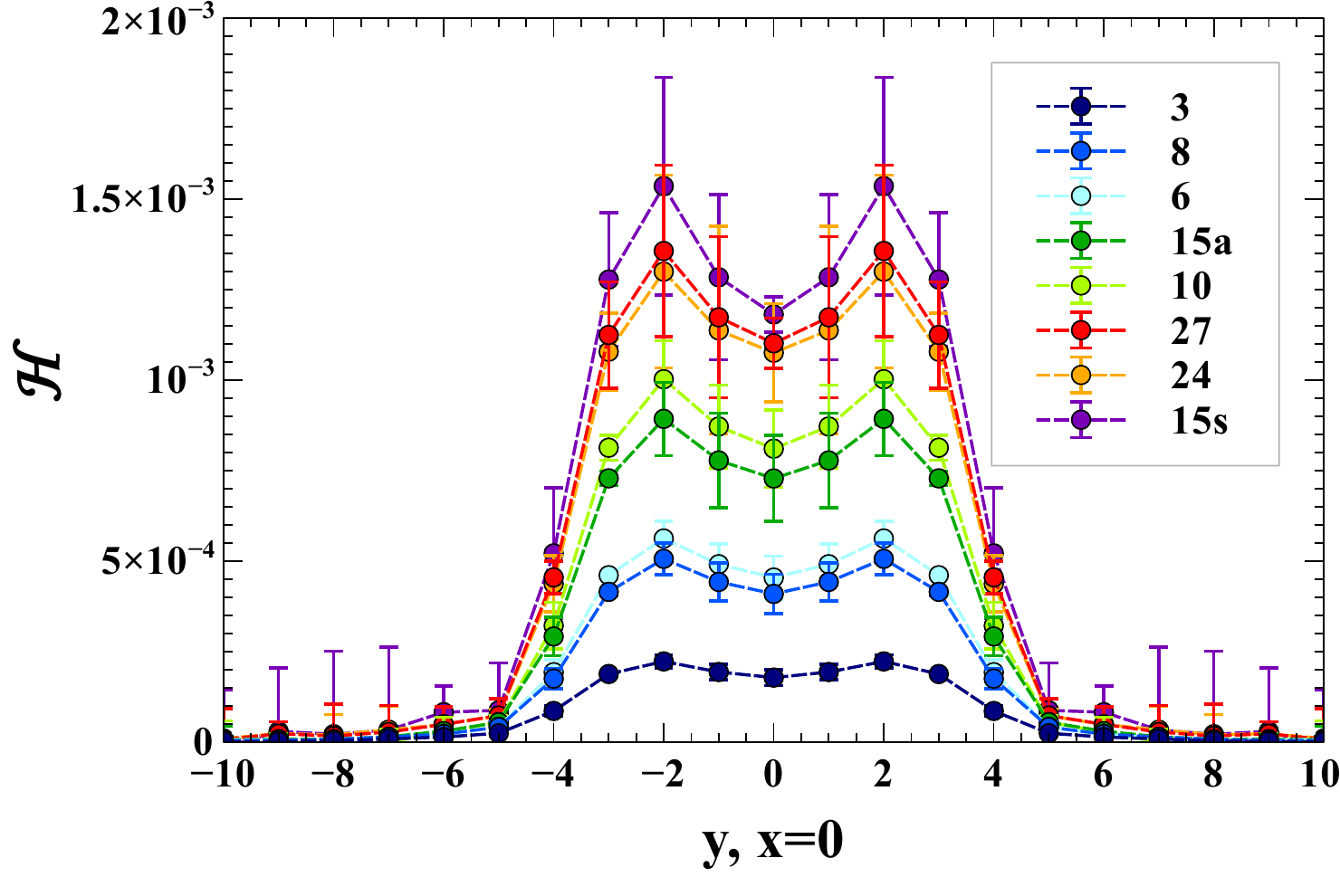}
    \includegraphics[width=0.4\linewidth]{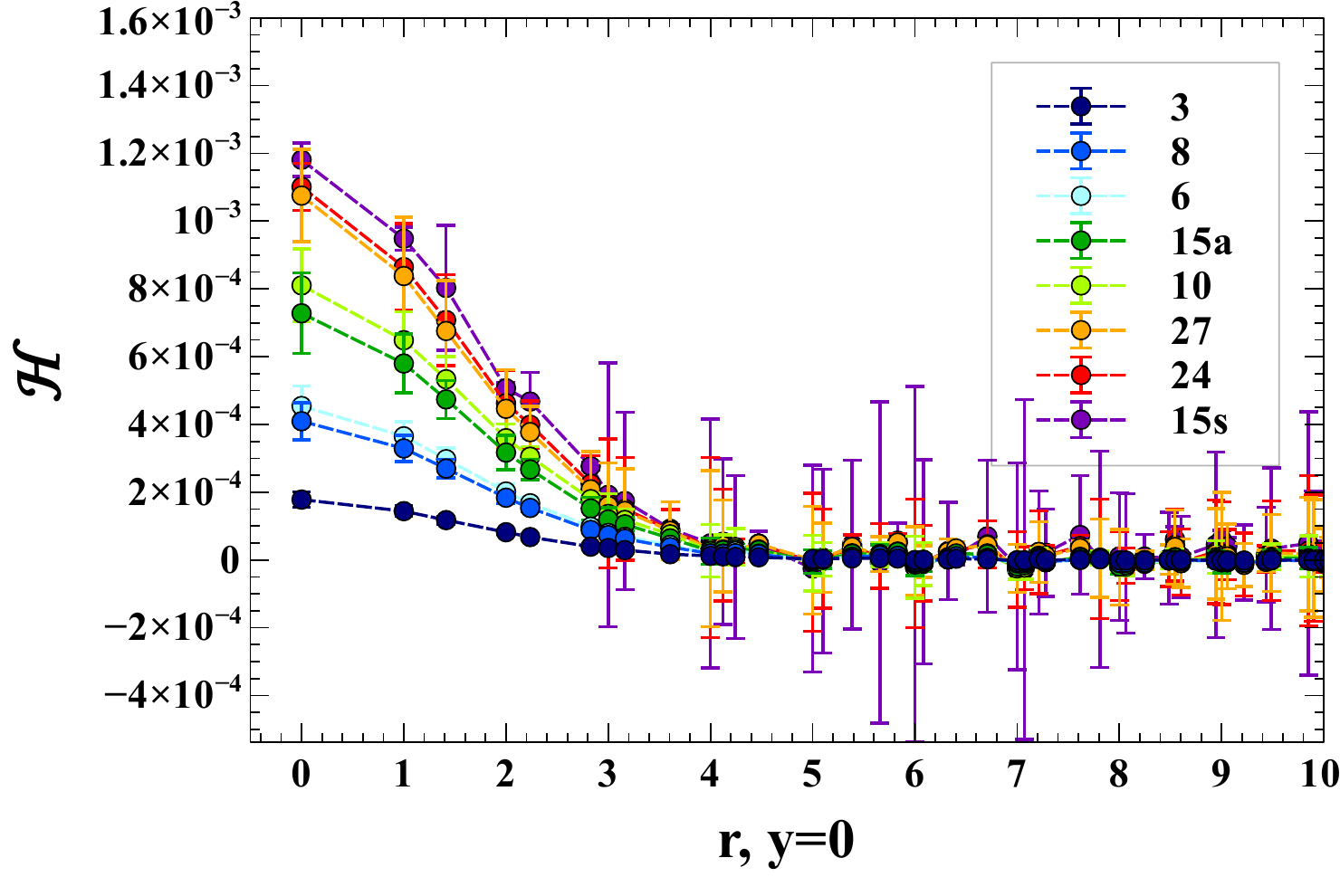}
    \caption{ Flux tube profiles for different representations of SU(3).
      On the left, results for  $x = 0$. On the right, for the mediator plane, with $r=\sqrt{x^2+z^2}$.}
    \label{multirep}
\end{figure}

In Fig. \ref{multirep} we present the flux tube profiles of the energy density for different representations of $SU(3)$.
In this case, we apply three levels of HYP in the temporal direction, in order
to further reduce the signal to noise ratio.

\begin{figure}[t!]
    \includegraphics[width=0.4\linewidth]{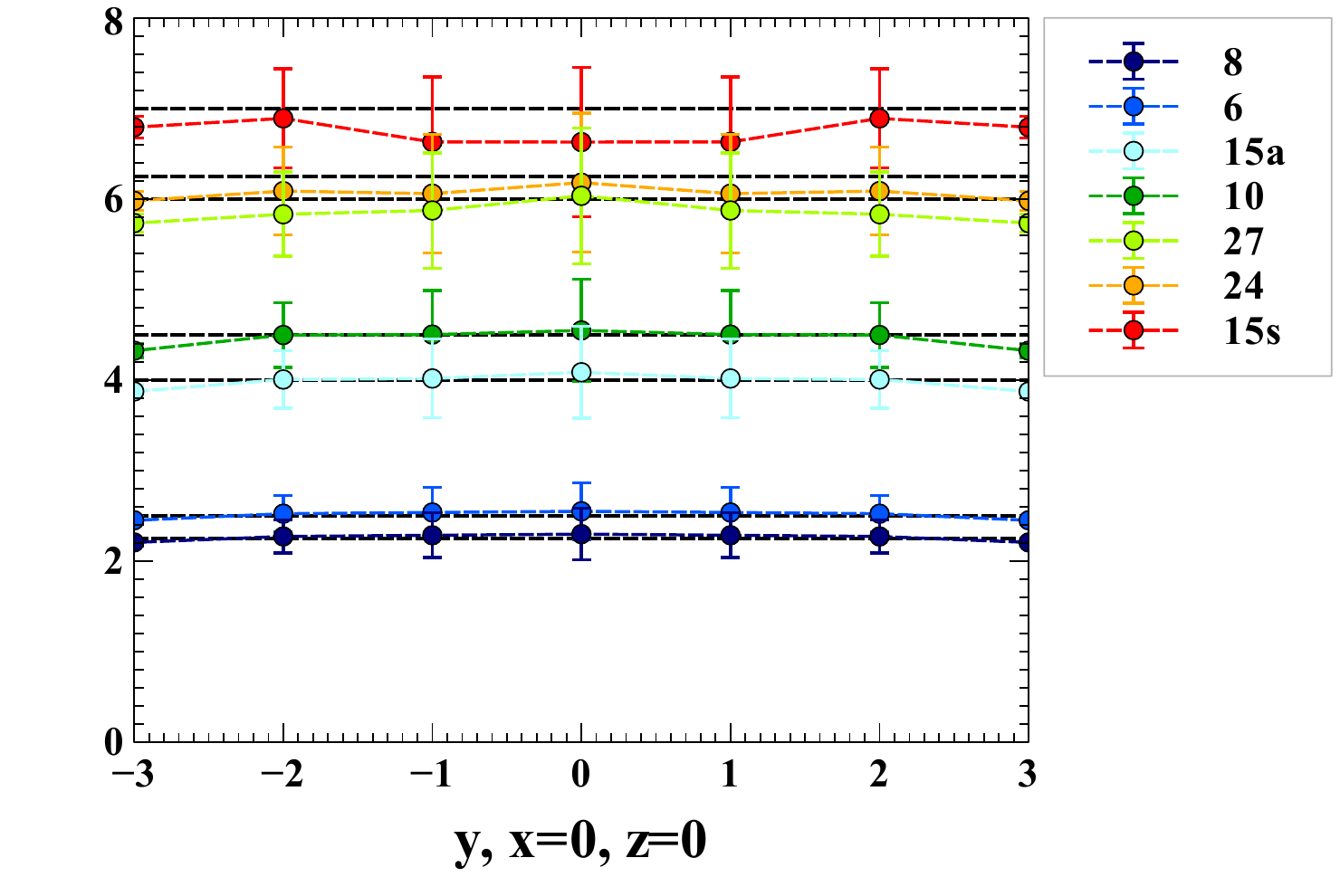}
    \includegraphics[width=0.4\linewidth]{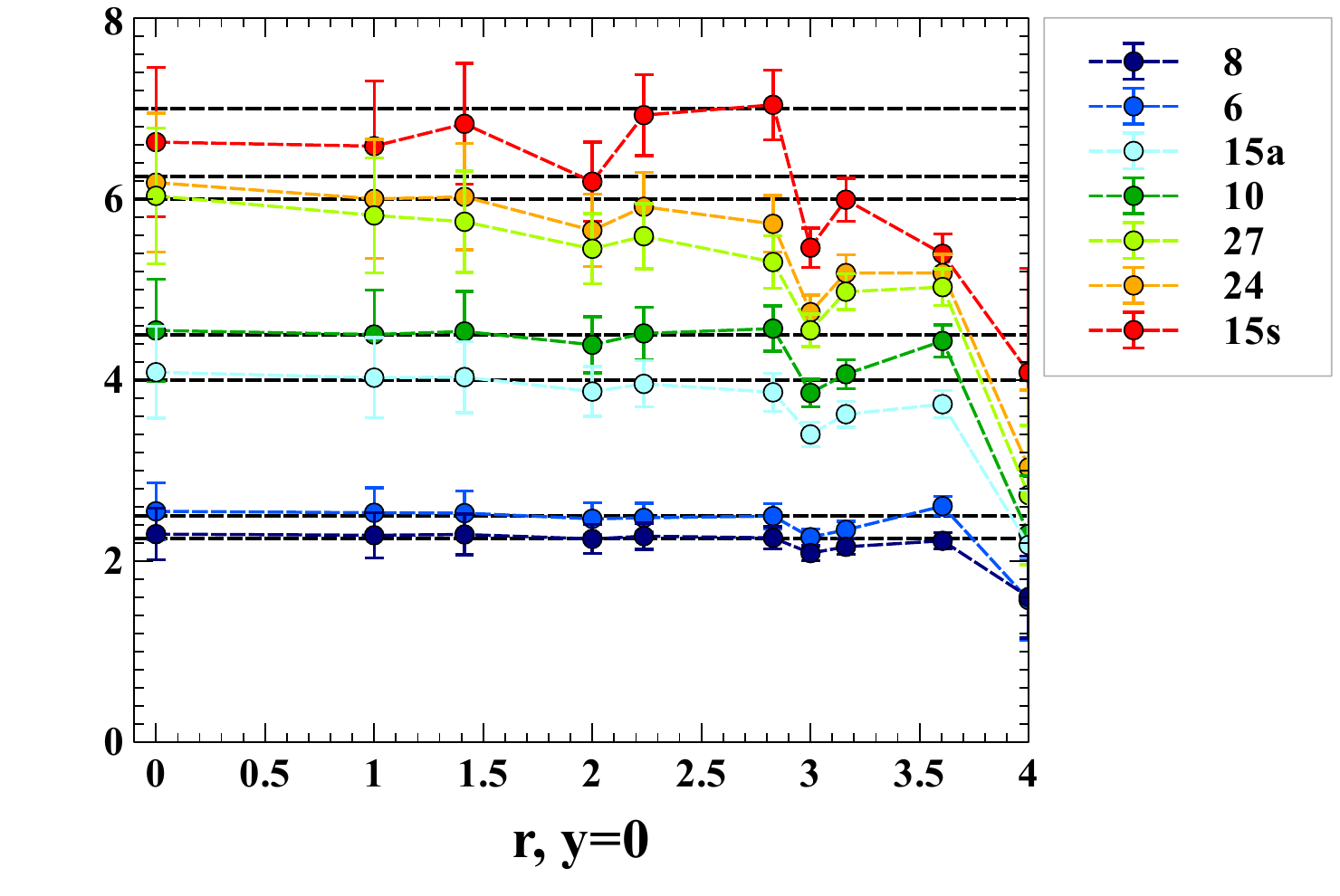}
    \caption{ 
    Ratios between the static potential of superior SU(3) representations to the fundamental one.
    The results are compared with the Casimir Scaling predictions ( black interrupted horizontal lines ).
    On the left the results are for $x = 0$. On the right they are for the mediator plane.
    }
    \label{casimir2}
\end{figure}

In Fig. \ref{casimir2} we see the ratios between the results for the flux tube profiles in the different
higher representations of $SU(3)$ and the fundamental one. The results presented are for short distancies, and seem to
agree with Casimir Scaling predictions. However, for larger distances the statistical errors, become greater
impeding us from concluding about the validity of the Casimir scaling hypothesis for the flux tube profiles.

\subsection{Dual Gluon Mass}

In 1970's, Nambu \cite{Nambu:1974zg}, 't Hooft \cite{'tHooft:1979uj} and Mandelstam \cite{Mandelstam:1974pi} proposed a very
interesting idea: quark confinement would be physically interpreted using the dual version of the superconductivity,
where the QCD vacuum state would behave like a magnetic superconductor.
 The chromoelectric field originated by a $q\overline{q}$ pair would be squeezed by a Meissner
effect into a dual Abrikosov flux tube, or even to QCD strings in the strong coupling limit, giving rise to the confining linear potential. 
Color confinement could be understood as the dual Meissner effect.

Notice that in common superconductivity the magnetic field decays with $B \sim e^{-r/\lambda_L}$
and this can be interpreted in terms of an effective mass for the photon $m_\gamma = 1 / \lambda_L$.
There is also evidence for the dual superconductor picture from numerical simulations or theoretical modles of QCD
\cite{Baker:1984qh,Bali:1996dm,Jia:2005sp}.

\begin{figure}[t!]
    \includegraphics[width=0.4\linewidth]{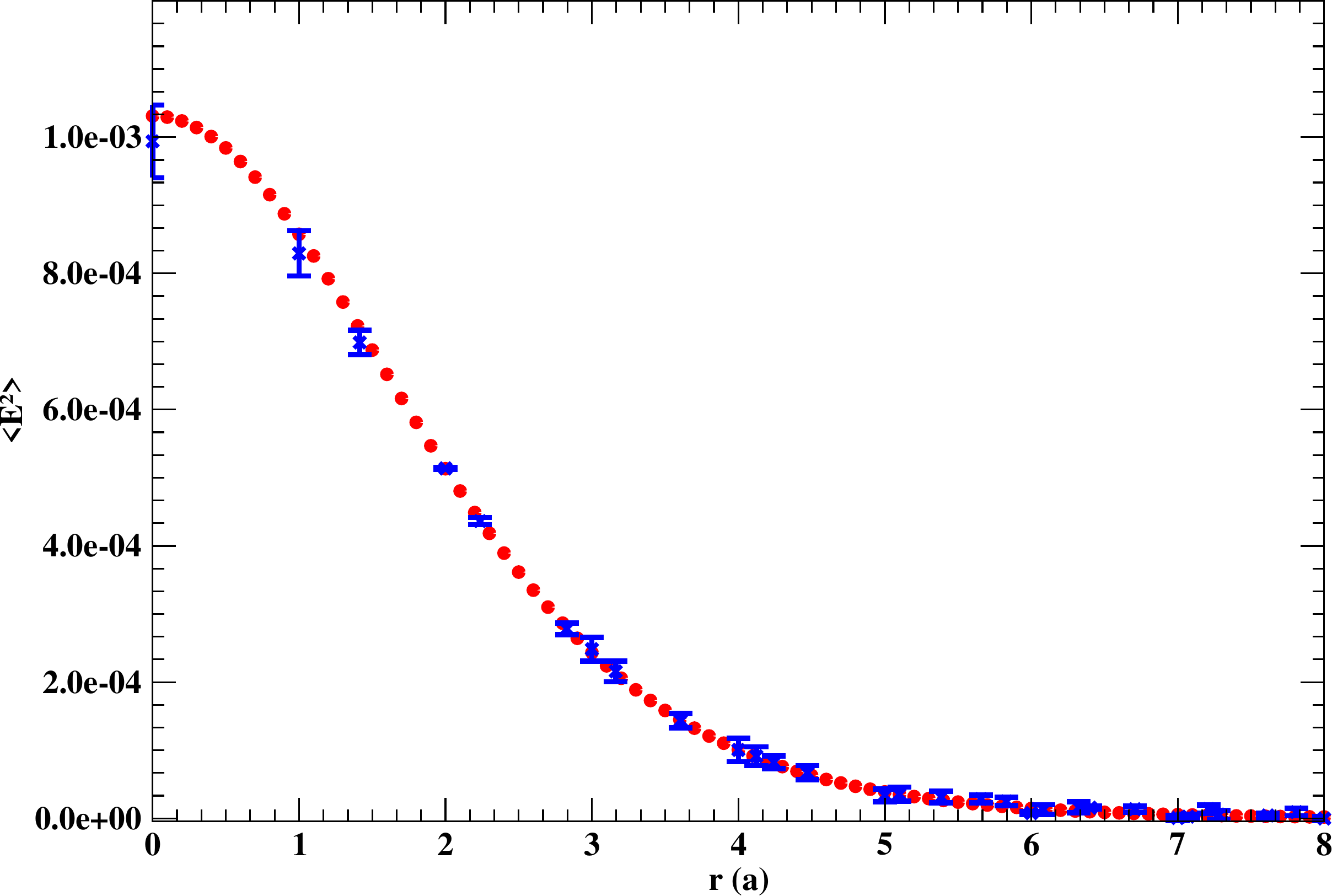}
    \includegraphics[width=0.4\linewidth]{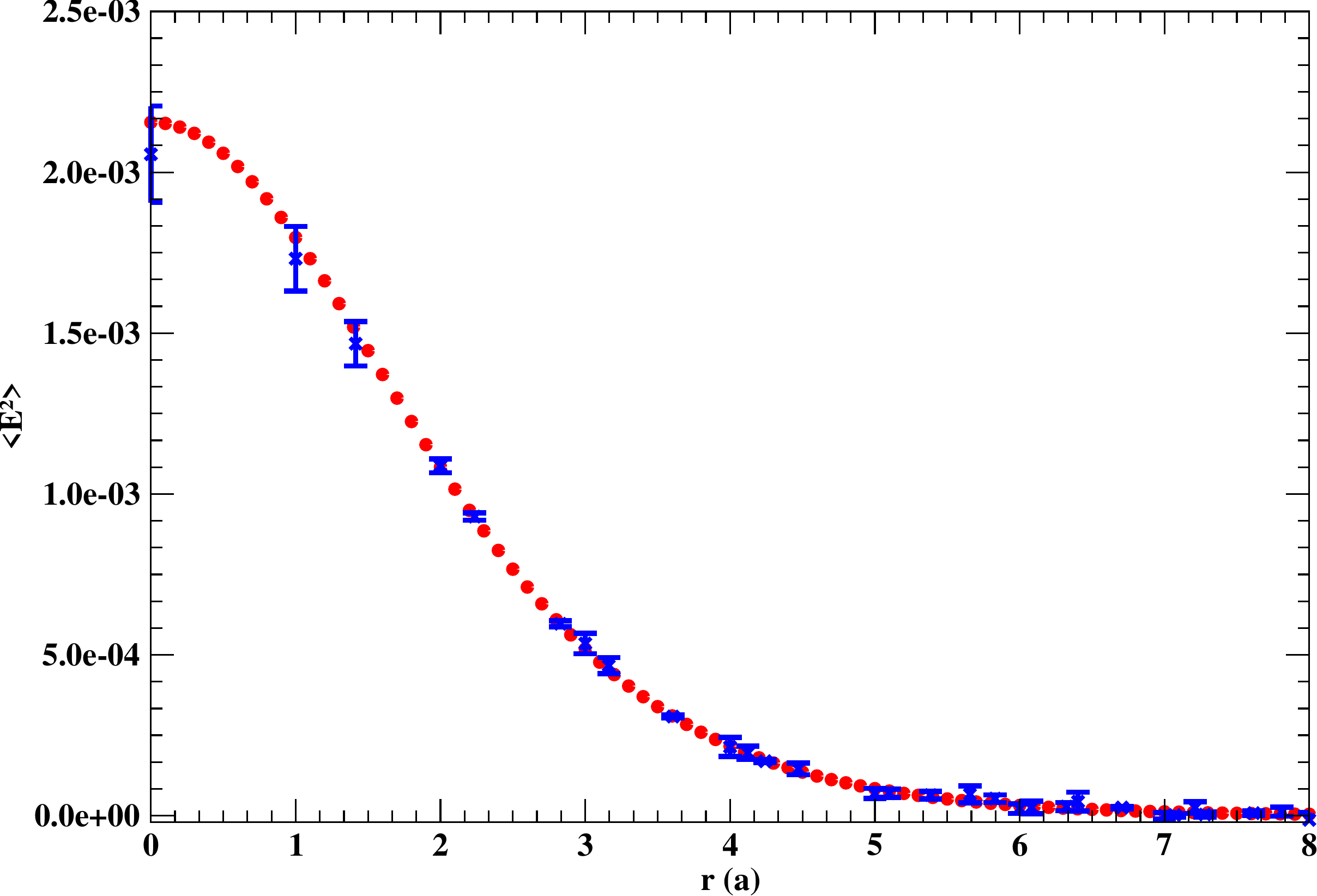}
    \caption{ 
    Fits of the profiles of the flux tubes of a meson (left) and of a glueball (right).
    }
    \label{profile fits}
\end{figure}

Here we test two functions, $a\, e^{-2\mu r}$ and $a\, K_0^2 \left(\mu r\right)$where $K_0$ is the modified Bessel function of order zero, and where $\mu=\frac{1}{\lambda_L}$, $\lambda_L$
is the penetration length.
So, in this case we have $\mu$ as a possible definition mass for the dual gluon.
Fitting the chromoelectric field and the lagrangian density section in the mid distance of the flux tube of the meson
and the two gluon glueball, we obtain the results presented in Table \ref{tab_dual_gluon_mass} for the effective
dual gluon mass, of the order of $ \sim 1\, GeV$.
Some values found in literature, for the effective dual gluon mass,
\cite{Suganuma:1997xk,Suganuma:2000jh,Suganuma:2002pm,Suganuma:2003ds}, and the effective gluon mass, \cite{Field:2001iu},
point to a similar value. Our fits are depicted in Fig. \ref{profile fits} and detailed
in Table \ref{tab_dual_gluon_mass}.

\section{Conclusions}

Notice that in type Type-II superconductors the flux tubes repel each other while
in Type-I superconductors they attract each other and tend to fuse in excited vortices. In QCD this raises the question, how the two pictures, of one adjoint string and of two fundamental strings, with different total string tensions,
match? 

By separating the quark from the antiquark in a static hybrid, we observe the fundamental string fusion and separation. 
When the quark and the anti-quark are superposed, we observe to the formation of an
adjoint string between the two gluon and agrees with Casimir Scaling measured by Bali \cite{Bali:2000un}.
This can be interpreted with a type-II superconductor analogy for the confinement in QCD with repulsion of the
fundamental strings and with the string tension of the first topological excitation of the string (the adjoint string)
larger than the double of the fundamental string tension.

We further study the flux tube created by higher representations of SU(3),
but above the decuplet 10 representation, it is not clear yet whether Casimir
scaling continues to occur in the field profiles of the flux tube. The detailed
study of the high representation field profiles still will continue to require
further and more detailed investigation.

We also compute a value for the dual gluon mass estimated form
the penetration length of the flux tube, to be $\lambda^{-1} = 0.91 \pm 0.16 $.
It is remarkable that this result is computed with gauge independent
lattice QCD calculations.

\begin{table}[t!]
\small{
\begin{centering}
\begin{tabular}{|c|c|c|c|c|}
\cline{2-5}
\multicolumn{1}{c|}{} & \multicolumn{2}{c|}{  $a\, e^{-2\, \mu\, r}$} & \multicolumn{2}{c|}{$a\, K_{0}^{2}\left(\mu\, r\right)$}\tabularnewline
\cline{2-5}
\multicolumn{1}{c|}{} & $\mu\ \left(GeV\right)$ &  $\chi^{2}/dof$ & $\mu\ \left(GeV\right)$ & $\chi^{2}/dof$\tabularnewline
\hline
\hline
$E_{(1)(a)}^{2}\left(r\right)$ & $1.170\pm0.228$ & $1.069$ & $0.805\pm0.287$ & $1.827$\tabularnewline
\hline
$\mathcal{L}_{(1)(a)}\left(r\right)$ & $1.170\pm0.119$ & $0.512$ & $0.865\pm0.188$ & $1.203$\tabularnewline
\hline
$E_{(2)(a)}^{2}\left(r\right)$ & $1.231\pm0.286$ & $1.547$ & $0.881\pm0.334$ & $2.084$\tabularnewline
\hline
$E_{(1)(b)}^{2}\left(r\right)$ & $1.210\pm0.056$ & $0.887$ & $0.897\pm0.085$ & $1.185$\tabularnewline
\hline
$\mathcal{L}_{(1)(b)}\left(r\right)$ & $1.208\pm0.068$ & $0.560$ & $0.909\pm0.099$ & $0.909$\tabularnewline
\hline
$E_{(2)(b)}^{2}\left(r\right)$ & $1.210\pm0.063$ & $1.162$ & $0.889\pm0.097$ & $1.262$\tabularnewline
\hline
$\mathcal{L}_{(2)(b)}\left(r\right)$ & $1.191\pm0.031$ & $1.066$ & $0.899\pm0.048$ & $1.106$\tabularnewline
\hline
\end{tabular}
\par\end{centering}
}
\caption{Results for the dual gluon mass, where (1) is for the two gluon glueball and (2) for the quark-antiquark cases,
and (a) at $y=4$ and $z=0$ with $r=x$ and (b) at $y=4$ with $r=(x,z)$.}
\label{tab_dual_gluon_mass}
\end{table}

\acknowledgments
We acknowledge discussions on superconductors with Pedro Sacramento.
We thank Orlando Oliveira for useful discussions and for sharing gauge field configurations.
This work was financed by the FCT contracts POCI/FP/81933/2007 and CERN/FP/83582/2008.

\end{document}